\documentclass[aps,prb,twocolumn,showpacs,superscriptaddress,amsmath,amssymb]{revtex4}

\usepackage{graphicx}
\usepackage{dcolumn}
\usepackage{bm}
\usepackage{color}
\RequirePackage[english]{babel}

\begin{document}

\title{Terahertz transition radiation of bulk and surface electromagnetic waves by an electron entering a layered superconductor }

\author{Yu.O.~Averkov}
\affiliation{ A.Ya.~Usikov Institute for Radiophysics and Electronics NASU, 61085 Kharkov, Ukraine}
\email{yuriyaverkov@gmail.com}

\author{V.M.~Yakovenko}
\affiliation{ A.Ya.~Usikov Institute for Radiophysics and Electronics NASU, 61085 Kharkov, Ukraine}
\email{yavm@ire.kharkov.ua}

\author{V.A.~Yampol'skii}
\affiliation{ A.Ya.~Usikov Institute for Radiophysics and Electronics NASU, 61085 Kharkov, Ukraine}
\affiliation{V.N.~Karazin Kharkov National University, 61077 Kharkov, Ukraine}
\affiliation{CEMS, RIKEN, Saitama 351-0198, Japan}
\email{yam@ire.kharkov.ua}

\author{Franco Nori}
\affiliation{CEMS, RIKEN, Saitama 351-0198, Japan}
\affiliation{University of Michigan, Ann Arbor, Michigan 48109, USA}
\email{fnori@riken.jp}

\begin{abstract}
We theoretically study the transition radiation of bulk and surface electromagnetic waves by an electron crossing an interface between a layered superconductor and an isotropic dielectric. We assume that the direction of the electron motion and the orientation of the superconducting layers are perpendicular to the interface. We derive the analytical expressions for the strongly anisotropic radiation fields and for the time-integrated energy fluxes of bulk and oblique surface electromagnetic waves (OSWs). We show that the OSWs with frequencies close to the Josephson plasma frequency $\omega_J$ provide the main contribution to the OSWs energy flux. Moreover, for frequencies close to the Josephson plasma frequency, the spectral density of the OSWs radiation diverges at some critical value of the azimuth angle $\varphi$. At the angles $\varphi=0$  and  $\varphi=90^o$, the radiation field has a transverse magnetic polarization. We have also studied the Cherenkov radiation by the electron escaping from the layered superconductor and show that this radiation is almost monochromatic. A remarkable feature of the Cherenkov radiation in a layered superconductor is that, contrary to the isotropic case, the Cherenkov radiation distinctly manifests itself in the angular dependence of the radiation energy flux.

\end{abstract}


\maketitle
\section{INTRODUCTION}

The transition radiation effect, i.e.~the radiation of electromagnetic waves by a uniformly moving electron due to its transition from one medium to another, was discovered by V.L. Ginzburg and I.M. Frank in 1945~\cite{1}. Since then, numerous monographs and reviews have been written on this subject (see, e.g., Refs.~\onlinecite{2,3,4,5,6}). Nowadays a good deal of attention to the transition radiation effect is caused by the large number of its applications. For instance, the transition radiation is used in high-energy physics for the detection of charged particles~\cite{2,Moran,Wakely}. The electron-bunch bombardment of solids and transit of the bunches through diaphragms  allow to generate short high-power pulses which are widely used for radiolocation~\cite{Buts2006}. The effect of transition radiation of anharmonic non-stationary electromagnetic pulses in free space and in a dispersive medium was investigated in Ref.~\onlinecite{10}. In particular, it was shown that the longitudinal profile of the bunch density and the plasma parameters can be determined from the values of the radiated fields and their derivatives near the leading edge of the resulting signal. The transition radiation of surface electromagnetic waves by a non-relativistic electron bunch which crosses the vacuum-semiconductor interface or a thin plate of a semiconductor  was studied  in Ref.~\onlinecite{11}, and now the radiation of modulated electron beams crossing a boundary of a plasma-like medium becomes a very effective method for generation of surface electromagnetic waves~\cite{12}.

The properties of the transition radiation by the electrons crossing anisotropic conducting interfaces, such as a wire shield, have been studied in Ref~\onlinecite{Aver2010}. Specifically, the possibility of obtaining an elliptical polarization of electromagnetic waves has been shown in this paper. V.E. Pafomov~\cite{17} predicted the Cherenkov radiation of the electromagnetic waves with negative group velocity  in media which possesse simultaneously negative permittivity and negative permeability  (so-called, left-handed media). A similar effect occurs if an electron crosses an interface between the vacuum and a uniaxial anisotropic conducting medium (e.g., a layered superconductor) in the case when \emph{the conducting layers are parallel to the interface}~\cite{18}. The electromagnetic waves in layered superconductors (so-called, Josephson plasma waves) belong to the terahertz frequency range, which is very important for various applications, but not easily accessible with modern electronic and optical devices. This technological perspective provides a strong motivation for studying these waves.

In this paper, we theoretically investigate the transition radiation produced by an electron entering (or escaping from) a layered superconductor \emph{with layers perpendicular to the interface}. We have studied in detail  the strongly-anisotropic angular distribution of the radiation energy flux and have shown that, in this geometry, unlike the results of Ref.~\onlinecite{18},  the excitation of the electromagnetic waves with negative group velocity is impossible. In other words, the directions of the electron motion and of the energy flux in the superconductor does not form an obtuse angle.  A remarkable feature of the Cherenkov radiation in a layered superconductor is that, contrary to the isotropic case, the Cherenkov radiation distinctly manifests itself in the angular dependence of the radiation energy flux. We have found that the electron transit from a dielectric to a layered superconductor results not only in the generation of \emph{bulk Josephson plasma waves}, but also in the excitation of \emph{oblique surface electromagnetic waves} (OSWs). We show that OSWs with frequencies close to the Josephson plasma frequency $\omega_J$ provide the main contribution to the OSWs energy flux. Moreover, for frequencies close to the Josephson plasma frequency, the spectral density of the OSWs radiation diverges at some critical value of the azimuth angle $\varphi$ with respect to the crystallographic \textbf{c}-axis.

\section{STATEMENT OF THE PROBLEM AND BASIC EQUATIONS}

\begin{figure}
\includegraphics [width=7.0 cm,height=4.5 cm]{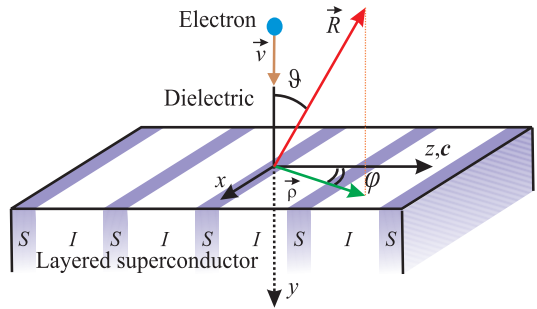}
\caption{\label{Fig1} (Color online) Geometry of the problem.}
\end{figure}

We consider a dielectric  with permittivity $\varepsilon_d$ and a layered superconductor with layers perpendicular to the interface (see Fig.~\ref{Fig1}). Both media are assumed to be nonmagnetic. We define our coordinate system so that the dielectric occupies the half-space $y<0$, the layered superconductor occupies the half-space $y>0$,  and the  $z$-axis coincides with the crystallographic $\bf{c}$-axis of the superconductor.

 Let an electron move uniformly in the dielectric along the $y$-axis with velocity $v\ll c$  (here $c$  is the speed of light in vacuum) and crosses the interface. The electron-charge density $n(\vec{r},t)$  is determined by the formula
\begin{equation}\label{Eq3}
    n(\vec{r},t)=e \delta (x)\delta (y-v t)\delta (z),
\end{equation}
where $\delta(x)$  is the Dirac delta function and $e$ is the electron charge.

We consider the Maxwell equations,
\begin{equation}\label{Eq4}
    \mathrm{rot}\vec{H}=\displaystyle\frac{\varepsilon(y)}{c}\frac{\partial \vec{E}}{\partial t}+\displaystyle\frac{4\pi}{c}\Bigl[\vec{j}_{\,\rm ind}(\vec{r},t)+\vec{j}_{\,\rm ext}(\vec{r},t)\Bigr],\
\end{equation}
\begin{equation}\label{Eq5}
    \mathrm{rot}\vec{E}=-\displaystyle\frac{1}{c}\frac{\partial \vec{H}}{\partial t}.\
\end{equation}
Here $\varepsilon(y)=\varepsilon_d $ for $y<0$,  $\varepsilon(y)=\varepsilon_s $ for $y>0$, $\varepsilon_s$ is the interlayer dielectric constant of the superconductor, $\vec{j}_{\,\rm ind}(\vec{r},t)$  is the current induced in the superconductor, and $\vec{j}_{\,\rm ext}(\vec{r},t)=e n(\vec{r},t) v$  is the electron current. The current $\vec{j}_{\,\rm ind}(\vec{r},t)$ inside the layered superconductor is determined by the distribution of the gauge-invariant phase difference $\phi(\vec{r},t)$ of the order parameter between neighboring layers. This phase difference is described by a set of coupled sine-Gordon equations (see, e.g., Ref.~\onlinecite{Thz-rev} and references therein). In the continuum and linear approximation, $\phi$ can be excluded from the set of equations for the electromagnetic fields, and the electrodynamics of layered superconductors can be described in terms of an anisotropic frequency-dependent permittivity with components $\varepsilon_c$ and $\varepsilon_{ab}$ across and along the layers, respectively. Indeed, as was shown in Ref.~\onlinecite{negref}, the direct solution of the set of coupled sine-Gordon equations in the continuum and linear approximations is equivalent to solving the Maxwell equations for a uniform anisotropic medium with
\begin{equation}\label{Eq1}
    \varepsilon_{ab}(\omega)=\varepsilon_s\Bigl(1-\displaystyle\frac{\gamma^2\omega_J^2}{\omega^2}+i\gamma^2\nu_{ab}\displaystyle\frac{\omega_J}{\omega}\Bigr),\
\end{equation}
\begin{equation}\label{Eq2}
    \varepsilon_{c}(\omega)=\varepsilon_s\Bigl(1-\displaystyle\frac{\omega_J^2}{\omega^2}+i\nu_{c}\displaystyle\frac{\omega_J}{\omega}\Bigr).
\end{equation}
Here $\omega_J=\bigl(8\pi e D J_c/\hbar \varepsilon_s\bigr)^{1/2}$ is the Josephson plasma frequency, $D$ is the period of the layered structure, $J_c$ is the maximal Josephson current density, $\gamma=\lambda_c/\lambda_{ab}\gg 1$ is the current-anisotropy parameter, $\lambda_c$ and $\lambda_{ab}$  are the magnetic-field penetration depths along and across the layers, respectively, the relaxation frequencies  $\nu_{ab}=4\pi\sigma_{ab}/(\varepsilon_s\omega_J\gamma^2)$ and $\nu_{c}=4\pi\sigma_{c}/(\varepsilon_s\omega_J)$  are proportional to the averaged quasiparticle conductivities $\sigma_{ab}$ (along the layers) and  $\sigma_c$ (across the layers), respectively.

\section{Electromagnetic field in the dielectric and in the layered superconductor}

The total electromagnetic field in each medium can be presented as a sum of the so-called electron's field and radiation field. The first of them represents a particular solution of the inhomogeneous Maxwell equations, while the latter is the solution of the homogeneous set of equations.  We express the electromagnetic fields in both media in terms of Fourier integrals over the coordinates $x$ and $z$, and over time $t$. For example, for the electric field $\vec{E}_{d,s}(\vec{r},t)$, we use the Fourier-series expansion,
\begin{equation}\label{Eq8}
    \vec{E}_{d,s}(\vec{r},t)=\int \vec{E}_{d,s}(\vec{\kappa},\omega,y) \exp\bigl[i (\vec{\kappa}\vec{\rho}-\omega t)\bigr] d\vec{\kappa} d\omega,
\end{equation}
where $\vec{\kappa}=(k_x, k_z)$ is the wave vector in the plane of the interface, $\vec{\rho}=(x,z)$ is the radius vector in the $xz$-plane, the subscripts $d$, $s$ denote the dielectric or superconductor regions ($y<0$ or $y>0$), respectively.

\subsection{Electron's electromagnetic field}

The Maxwell equations give the following expressions for the electron's electric $\vec{E}_d^{({\rm el})}(\vec{\kappa},\omega,y)$ and magnetic $\vec{H}_d^{({\rm el})}(\vec{\kappa},\omega,y)$  fields in the dielectric:
\begin{equation}\label{Eq9}
    \vec{E}_d^{({\rm el})} (\vec{\kappa},\omega,y)=\displaystyle\frac{i e}{2\pi^2 v\, \varepsilon_d}\frac{\displaystyle\frac{\omega\vec{v}}{c^2}\varepsilon_d -\vec{k}}{k^2-\displaystyle\frac{\omega^2}{c^2} \varepsilon_d}\exp\left(i\frac{\omega}{v} y\right),\
\end{equation}
\begin{equation}\label{Eq10}
    \vec{H}_d^{({\rm el})} (\vec{\kappa},\omega,y) =\displaystyle\frac{\varepsilon_d}{c}\Bigl[\vec{v}\times\vec{E}_d^{({\rm el})} (\vec{\kappa},\omega,y)\Bigr].
\end{equation}
Here the wave vector $\vec{k}$ has the components ($k_x$, $k_y=\omega/v$, $k_z$), $k=|\vec{k}|$.

The electron's field components in the superconductor are presented in Appendix A.
In the next subsection, we consider the second component of the electromagnetic field, namely, the radiation field.

\subsection{Radiation field}

 It is suitable to present the radiation field as a sum of ordinary and extraordinary electromagnetic waves with components $(E^{(\rm ord)}_{d,s\,x}, E^{(\rm ord)}_{d,s\,y}, 0)$, $(H^{(\rm ord)}_{d,s\,x},H^{(\rm ord)}_{d,s\,y},
 H^{(\rm ord)}_{d,s\,z})$ and $(E^{(\rm ext)}_{d,s\,x}, E^{(\rm ext)}_{d,s\,y}, E^{(\rm ext)}_{d,s\,z})$, $(H^{(\rm ext)}_{d,s\,x}, H^{(\rm ext)}_{d,s\,y}, 0)$, respectively.
The expressions for the Fourier components of the electric and magnetic fields in the ordinary and extraordinary waves are presented in Appendix B.

\subsection{Total radiation field in the dielectric}

From the continuity conditions for the tangential components of the electric, [$E_x(y=0)$, $E_z(y=0)$],  and magnetic, [$H_x(y=0)$, $H_z(y=0)$], fields at the interface, we derive the expressions for four unknown Fourier amplitudes,  $E^{({\rm ord})}_{d\,x}(\vec{\kappa},\omega,y)$, $E^{({\rm ord})}_{s\,x}(\vec{\kappa},\omega,y)$, $E^{({\rm ext})}_{d\,x}(\vec{\kappa},\omega,y)$, and $E^{({\rm ext})}_{s\,x}(\vec{\kappa},\omega,y)$. Since the ordinary wave in the layered superconductor is evanescent for frequencies $\omega \ll \omega_J \gamma$, the \emph{propagating} field with $|\mathrm{Re}k_{s\,y}|\gg |\mathrm{Im}k_{s\,y}|$ in the superconductor is represented by the extraordinary wave only. It is difficult to observe the radiation field inside the layered superconductor, and, therefore, below we analyze the electromagnetic fields only in the dielectric (e.g., in the vacuum). The expressions for all components of the total radiation field in the dielectric are presented in Appendix C.

In order to derive the spatial and temporal dependences of the total electromagnetic fields in explicit form, we need to integrate expressions (C1--C6) in Appendix C over $k_x$  and $k_z$  by means of the stationary-phase method for double integrals~\cite{19}. Using this method, we obtain the following quite clear result for the stationary points:
\begin{equation}\label{Eq51}
    k_{x0}=\displaystyle\frac{\omega}{c}\sqrt{\varepsilon_d}\sin\vartheta\sin\varphi,\
\end{equation}
\begin{equation}\label{Eq52}
    k_{z0}=\displaystyle\frac{\omega}{c}\sqrt{\varepsilon_d}\sin\vartheta\cos\varphi,\
\end{equation}
where $\vartheta$  is the tilt angle with respect to the  $y$-axis (i.e., the angle between the radius vector $\vec{R}$ and the $y$-axis), $\varphi$  is the azimuth angle in the interface plane (i.e., the angle between the radius vector $\vec{\rho}$ and the $z$-axis)  (see Fig.~\ref{Fig1}). The stationary-phase method is applicable only for the long distances $R$ between the point where the electron crosses the interface and the observation point; namely, when the following condition holds true \cite{Garibian_1957}:
 \begin{equation}\label{FormZone}
    R\gg \displaystyle\frac{c}{\omega \sin^2\vartheta}.
 \end{equation}
In what follows, we assume that this condition is satisfied.

To calculate the energy losses of the electron due to radiation into the dielectric, we find the energy flux of the total electromagnetic wave in the dielectric across the remote hemisphere using the time-integrated Poynting vector,
\begin{equation}\label{Eq53}
    \langle \vec{S} \rangle=\displaystyle\frac{c}{4\pi}\mathrm{ Re}\int\limits_{-\infty}^{\infty} dt \Bigl[\vec{E}(\vec{r},t),\vec{H}^*(\vec{r},t)\Bigr].
\end{equation}
Eventually, we obtain the following expression for the spectral density $\Pi (\Omega,\vartheta,\varphi)$  of the radiation energy flux (in units of $\Pi_0=e^2/c$) per unit solid angle $\Theta$ ($d\Theta=\sin\vartheta d\vartheta d\varphi$) integrated over the electron transit time:
\begin{eqnarray}
 &   \Pi (\Omega,\vartheta,\varphi)=2\pi^2 \varepsilon_d \Omega^2 \cos^2\vartheta\nonumber\\
&\times \mathrm{Re}\Biggl\{\Bigl[\bar{E}_{d\,y}(\Omega,\vartheta,\varphi) \bar{H}_{d\,z}^*(\Omega,\vartheta,\varphi)\nonumber\\
&- \bar{E}_{d\,z}(\Omega,\vartheta,\varphi) \bar{H}_{d\,y}^*(\Omega,\vartheta,\varphi)\Bigr]\sin\vartheta\sin\varphi \nonumber\\
& +\Bigl[\bar{E}_{d\,x}(\Omega,\vartheta,\varphi)\bar{H}_{d\,y}^*(\Omega,\vartheta,\varphi) \nonumber\\
& - \bar{E}_{d\,y}(\Omega,\vartheta,\varphi) \bar{H}_{d\,x}^*(\Omega,\vartheta,\varphi)  \Bigr]\sin\vartheta\cos\varphi \nonumber\\
&- \Bigl[\bar{E}_{d\,z}(\Omega,\vartheta,\varphi) \bar{H}_{d\,x}^*(\Omega,\vartheta,\varphi) \nonumber\\
&  -  \bar{E}_{d\,x}(\Omega,\vartheta,\varphi) \bar{H}_{d\,z}^*(\Omega,\vartheta,\varphi)\Bigr]\cos\vartheta \Biggr\}
\end{eqnarray}
where $\bar{E}_{d\,i}(\Omega,\vartheta,\varphi)$ and $\bar{H}_{d\,i}(\Omega,\vartheta,\varphi)$  are dimensionless Fourier components for the radiation fields (in units of $e/\omega_J$) given by equations (C1--C6) in Appendix C and expressed in terms of the dimensionless frequency $\Omega = \omega/\omega_J$  taking into account Eqs.~\eqref{Eq51} and \eqref{Eq52}. It is important to emphasize that, for the distances $R$ satisfying inequality \eqref{FormZone}, the radiated waves lose touch with the superconductor, and their spectral density $\Pi (\Omega,\vartheta,\varphi)$ does not depend on $R$, even when accounting for the losses in the superconductor.

\section{ANALYSIS OF THE RADIATION SPECTRUM OF BULK ELECTROMAGNETIC WAVES}

\begin{figure}
\includegraphics [width=7.0 cm,height=6 cm]{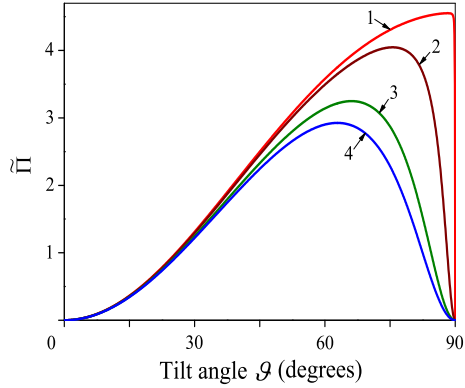}
\caption{\label{Fig2} (Color online)  Dependence of the dimensionless spectral density $\tilde{\Pi}=10^3\,\Pi \cdot c/e^2$ of the radiation on the tilt angle $\vartheta$ for different values of $\varphi$, at $\Omega=0.7$ and $\beta=0.3$. Curves 1--4 correspond to  $\varphi=90^o$, $60^o$,  $30^o$,  and $\varphi=0$, respectively.}
\end{figure}

In this section, we perform the numerical analysis of the dependence of the spectral density $\Pi (\Omega,\vartheta,\varphi)$  on the tilt angle  $\vartheta$ and the azimuth angle $\varphi$.  Hereafter, we use the following parameters for the adjacent media:
\begin{equation}\label{Eq55}
    \varepsilon_d=1,\ \varepsilon_s=16,\ \gamma=200,\ \nu_{ab}=0,\ \nu_c=10^{-5}.
\end{equation}
Figure~\ref{Fig2} shows the dependence of  $\Pi$ on the tilt angle $\vartheta$  for different values of $\varphi$,  at $\Omega=0.7$ and $\beta=v/c=0.3$,  for the case when the electron enters from the dielectric into the superconductor. As seen from Fig.~\ref{Fig2}, the maximum of the spectral density is located at $\vartheta\approx 90^o$ and $\varphi=90^o$.
In other words,  for $\varphi\approx 90^o$, the radiation energy flux is directed at a grazing angle with respect to the interface. It is also seen from Fig.~\ref{Fig2} that the maximum of the spectral density decreases and shifts towards lower angles $\vartheta$  when decreasing the angle $\varphi$.  Note that the radiation field has the transverse magnetic (TM) polarization at angles $\varphi\approx 0$ and  $\varphi\approx  90^o$. At $\varphi\approx 0^o$,  the radiation field has components  $(0, E_{d\,y}, E_{d\,z})$,  $(H_{d\,x},0,0)$, while at $\varphi\approx 90^o$  it  has components $(E_{d\,x}, E_{d\,y},0)$,  $(0,0,H_{d\,z})$.

\begin{figure}
\includegraphics [width=7.0 cm,height=6 cm]{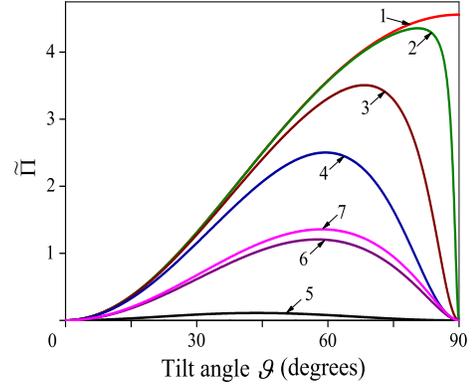}
\caption{\label{Fig3} (Color online) Dependence of the dimensionless spectral density $\tilde{\Pi}=10^3\,\Pi \cdot c/e^2$  of the radiation on the tilt angle $\vartheta$  for a number of frequencies $\Omega$, at $\varphi=0$ and $\beta=0.3$. Curves 1--7 correspond to $\Omega= 10^{-3}$,  $0.1$, $0.5$, $0.8$,  $1$,   $2$,  and $10$, respectively. }
\end{figure}

\begin{figure}
\includegraphics [width=7.0 cm,height=6 cm]{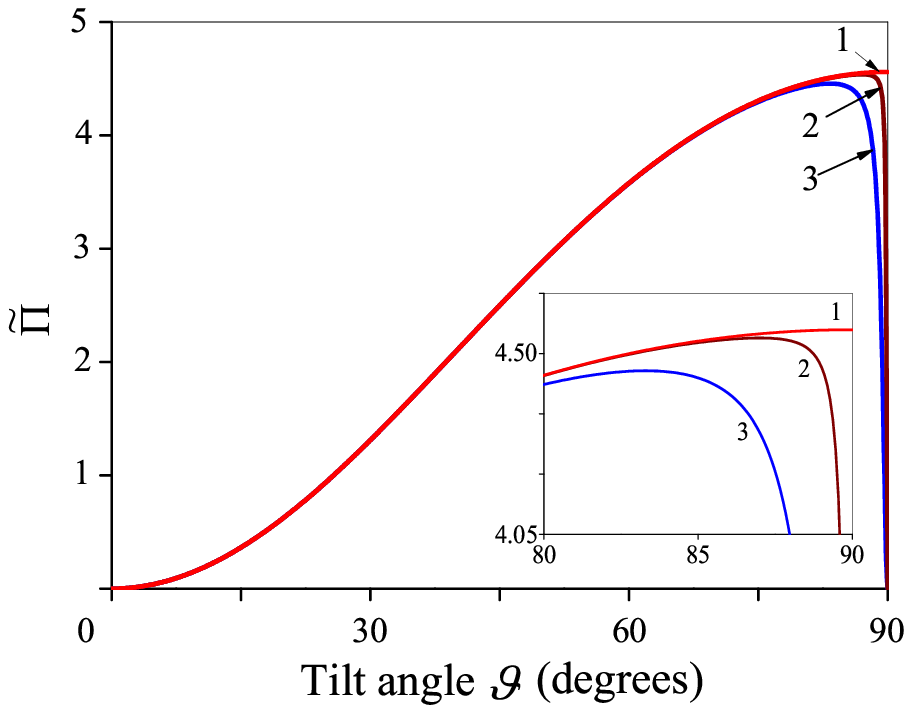}
\caption{\label{Fig4} (Color online) Dependence of the dimensionless spectral density $\tilde{\Pi}=10^3\,\Pi \cdot c/e^2$   of the radiation on the tilt angle $\vartheta$  for a number of frequencies $\Omega$, at $\varphi=90^o$ and $\beta=0.3$. Curves 1, 2, 3 correspond to $\Omega= 10^{-3}$,  $2$,  and $10$, respectively.}
\end{figure}

The dependence of $\Pi $ on the tilt angle $\vartheta$ for a number of frequencies $\Omega$  at  $\varphi=0$ and $\beta=0.3$  is shown in Fig.~\ref{Fig3}.
One can see that, for small frequencies, the maximum of the spectral density is located at grazing angles $\vartheta\approx 90^o$.  When the frequency increases, the maximum of the spectral density changes non-monotonically. At first, it decreases and tends to a certain minimum value at $\Omega=1$  (curve 5), but then the maximum grows and tends to some limit value at $\Omega\gg 1$  (curve 7). Fig.~\ref{Fig4} presents the analogous dependence for $\varphi=90^o$ and $\beta=0.3$.
The curves in this figure demonstrate that, for all frequencies which have a physical meaning, the maxima of the spectral density are observed for grazing tilt angles $\vartheta$ close to $90^o$.

\subsection{Cherenkov radiation by the electron escaping from the layered superconductor}

It is worthwhile to consider the Cherenkov radiation generated by the electron in the layered superconductor. Note first that, according to Eq.~(B10) in Appendix B,  the $y$-component of the wave vector of ordinary Josephson plasma waves is imaginary for all reasonable frequencies. This means that ordinary waves cannot propagate in the superconductor along the $y$-axis: they are evanescent modes. Therefore, Cherenkov radiation can be presented in the layered superconductor by the extraordinary waves only. On the other hand, as was shown in Refs.~\onlinecite{negref,neg-ref},  the $y$-components of the wave vector and the Poynting vector of an extraordinary Josephson plasma wave have the same sign. In other words, for the geometry considered in this paper, the Cherenkov radiation is oriented in the same direction as the velocity of the moving electron, contrary to the geometry with layers parallel to the interface, considered in Ref.~\onlinecite{18}. Such a result was obtained for the Cherenkov radiation produced by a Josephson vortex moving along the superconducting layers~\cite{vortex}. Hence, the orientation of the superconducting layers with respect to the interface plane plays a crucial role for the formation of the reversed or non-reversed Cherenkov radiation. This means that, in our geometry, the Cherenkov radiation can escape from the superconductor only in the case when the electron moves towards the interface. Therefore, in this section, we consider the case of negative electron velocity which corresponds to the electron escaping from the superconductor.

As is well-known, the Cherenkov radiation can arise if the electron velocity $v$ coincides with the $y$-component of the wave phase velocity,  $v=\omega/ k_{s\,y}^{({\rm ext})}$. A simple analysis shows that this condition, with  $k_{s\,y}^{({\rm ext})}$ given by Eq.~(B19) in Appendix B, can be satisfied only in a narrow frequency interval near the frequency  $\omega_{_{\rm Ch}}$,
\begin{equation}\label{Eq61}
    \omega_{_{\rm Ch}}=\omega_J\Bigl[1-\bigl(\varepsilon_s\beta^2\bigr)^{-1}\Bigr]^{-1/2}, \quad \beta=\frac{v}{c},
\end{equation}
if the permittivity $\varepsilon_d$ of the dielectric is small with respect to the interlayer permittivity $\varepsilon_s$ of the superconductor. Here we assume that $\beta^2 > 1/\varepsilon_s$.  In other words, the Cherenkov radiation appears to be almost monochromatic.

\begin{figure}
\includegraphics [width=7.0 cm,height=6 cm]{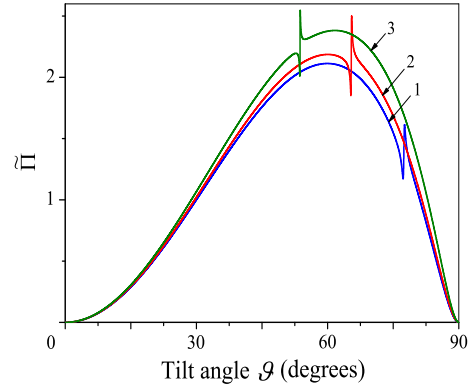}
\caption{\label{Fig5} (Color online) Dependence of the dimensionless spectral density $\tilde{\Pi}=10^3\,\Pi \cdot c/e^2$  of the radiation on the tilt angle $\vartheta$ for different values of $\varphi$ at $\Omega=1.87$ and $\beta=-0.3$. Curves 1, 2, 3 correspond to $\varphi=35^o$, $38^o$,  and $44^o$, respectively}.
\end{figure}

Figure~\ref{Fig5} shows the dependence of the spectral density $\Pi $  of the radiation on the tilt angle $\vartheta$ for several small values of the azimuth angle $\varphi$ at $\Omega=1.87$, $\nu_{c}=5\cdot 10^{-5}$,  and $\beta=-0.3$. The sharp maxima in the curves correspond to the Cherenkov radiation. It is important to note that, in the isotropic case, the input of the Cherenkov radiation to the spectral  density $\Pi (\vartheta)$ cannot be distinguished on the background of the spectral density of the total transition radiation~\cite{3}. However, as seen in Fig.~\ref{Fig5}, the input of the Cherenkov radiation to $\Pi (\vartheta)$ distinctly manifests itself in the anisotropic case. This is a remarkable new feature of the Cherenkov radiation in anisotropic media.

\section{Transition radiation  OF THE OBLIQUE SURFACE ELECTROMAGNETIC WAVES}

It is important to emphasize that the moving electron excites not only the above-considered bulk electromagnetic waves, but also the so-called oblique surface electromagnetic waves (OSWs) in the interface plane. The properties of the OSWs have recently been investigated in Refs.~\onlinecite{20,21}. Here we consider the excitation of the OSWs due to the transition radiation by the moving electron.

The factor $\Delta_{\rm OSW}$ in the denominators of Eqs.~(C1--C6) in Appendix C defines  the dispersion relation for the OSWs, $\Delta_{\rm OSW}=0$. The results for the OSWs fields calculation are presented in Appendix D.

In order to calculate the integrated density $\Pi_{\rm OSW}$ of the OSWs energy flux over the electron transit time, we use Eq.~\eqref{Eq53}. Since the OSWs are evanescent waves, the OSWs Poynting vector is directed strictly along the interface, and the features of the OSWs excitation cannot be seen in the spectral density of the waves emitted into the upper hemisphere. Therefore, to find the OSWs energy flux over the electron transit time we should integrate $\langle S_{\rho} \rangle=\langle S_x \rangle\sin\varphi+\langle S_z \rangle\cos\varphi$ over the lateral surface of the semi-infinite circular cylinder  with radius $\rho$ and with the axis parallel to the electron path in the half-space $y<0$. For calculation of $\langle S_{\rho} \rangle$, we evaluate the integrals (D1--D6) in Appendix D over $\psi$ using the stationary-phase method~\cite{22}. For the case of OSWs radiation, the condition \eqref{FormZone} of the method applicability should be replaced by $\rho\gg 1/\kappa_0(\omega,\psi)$, where $\kappa_0(\omega,\psi)$ is the solution of the dispersion relation $\Delta_{\rm OSW}=0$ (see Appendix D).
Here we analyze the dependence of $\Pi_{\rm OSW}$ [normalized to $\pi^3 e^2/(2 c)$] on the azimuth angle $\varphi$ and on the dimensionless frequency $\Omega$. It is worthwhile to note that, for the calculation of integrals over the angle $\psi$ by the stationary phase method, it is useful to use the asymptotic solution of the dispersion relation $\Delta_{\rm OSW}=0$ for the OSWs,
 \begin{equation}\label{EqDisp}
    \bar{\kappa}=\displaystyle\frac{\Omega}{\sqrt{2}\sin\psi}\sqrt{\zeta+\sqrt{\zeta^2-4 \varepsilon_d (\varepsilon_c-\varepsilon_d)\tan^2\psi}},
 \end{equation}
 where $\bar{\kappa}=c \kappa /\omega_J$ and $\zeta=\varepsilon_c(\omega)+[\varepsilon_c(\omega)-\varepsilon_d]\tan^2\psi$. This asymptotic formula properly describes the dispersion relation for $0\leq\psi\leq \pi/2$, $0<\Omega<\Omega_1=\sqrt{\varepsilon_s/(\varepsilon_s-\varepsilon_d)}$, when the following inequalities are satisfied:
 \begin{equation}\label{EqCond}
    \Omega^2 |\varepsilon_{ab}(\omega)|\gg \bar{\kappa}^2,\ \qquad    \Biggl|\displaystyle\frac{\varepsilon_c(\omega)}{\varepsilon_{ab}(\omega)}\Biggr|\ll 1.
 \end{equation}
 Here the frequency $\Omega_1$ is the frequency for the end-points of the OSWs dispersion curves \cite{20}.

\begin{figure}
\includegraphics [width=8 cm,height=6 cm]{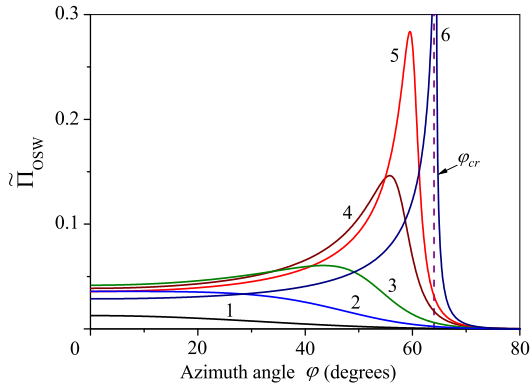}
\caption{\label{Fig6} (Color online) Dependence of the dimensionless spectral density $\tilde{\Pi}_{\rm OSW}=\Pi_{\rm OSW} \cdot 2c/\pi^3e^2$ of the OSW radiation on the azimuth angle $\varphi$ for a number of frequencies $\Omega$ at $\varepsilon_d=1$ and $\nu_{ab}=\nu_c=0$. Curves 1--6 correspond to $\Omega=0.5$, $0.9$, $0.96$, $0.98$, $0.985$, and $0.991$, respectively. The frequency $\Omega=0.991$ matches the critical azimuth angle $\varphi_{\rm cr}\approx 63.63^o$.}
\end{figure}

Figure~\ref{Fig6} demonstrates the dependence of the spectral density $\Pi_{\rm OSW}$ of the OSW radiation on the azimuth angle $\varphi$ for a number of frequencies $\Omega$ without regard to losses in the superconductor (i.e., for  $\nu_{ab}=\nu_c=0$) at $\varepsilon_d=1$.
Curves 1 and 2 in this figure  show that, for relatively low frequencies, the $\Pi_{\rm OSW}(\varphi) $ dependence is too weak and does not contain maxima.  As the frequency increases, pronounced maxima appear (see curves 3, 4, and 5). Moreover, at frequencies higher than some critical frequency $\Omega_{\rm cr}\approx 0.991$,  the dependence $\Pi_{\rm OSW} (\varphi)$ has a divergence at a certain value $\varphi_{\rm cr}$ of the azimuth angle. This is demonstrated by curve 6, for which $\varphi_{\rm cr}\approx 63.63^o$. As the frequency $\Omega$ tends to one, the divergence occurs at larger angles $\varphi_{\rm cr}$, and the critical value of the azimuth angle tends to $90^o$ in the limit $\Omega\rightarrow 1$. Note that, in the isotropic case, when the surface electromagnetic waves are radiated by an electron crossing the interface between the vacuum and an isotropic plasma-like medium, the spectral density has a divergence {\it at the frequency of surface electrostatic oscillations}~\cite{3}, and this divergence occurs at the same frequency for any value of the azimuth angle $\varphi$. Thus, the anisotropy of the interface causes the dependence of the critical frequency $\Omega_{\rm cr}$ on the azimuth angle $\varphi$.

\begin{figure}
\includegraphics [width=14 cm,height=6 cm]{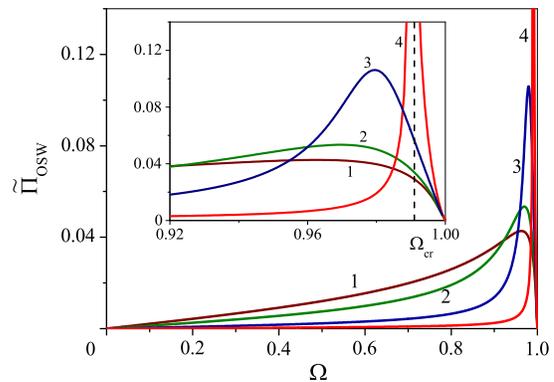}
\caption{\label{Fig7} (Color online) Frequency dependence of the dimensionless OSWs spectral density $\tilde{\Pi}_{\rm OSW}=\Pi_{\rm OSW} \cdot 2c/\pi^3e^2$ for a number of values of the azimuth angle $\varphi$, at $\varepsilon_d=1$ and $\nu_{ab}=\nu_c=0$. Curves 1--4 correspond to $\varphi=10^o$, $30^o$, $50^o$,  and $\varphi\approx 63.63^o$. The angle $\varphi\approx 63.63^o$ matches the critical frequency $\Omega_{cr}\approx 0.991$, which is shown by the dashed line in the inset.}
\end{figure}

The frequency dependence of the OSWs spectral density $\Pi_{\rm OSW}$ of the OSW radiation for a number of values of $\varphi$ without regard to losses in the superconductor (i.e., for $\nu_{ab}=\nu_c=0$) and at $\varepsilon_d=1$ are presented in Fig.~\ref{Fig7}.
 As seen in this figure, the OSWs spectral density increases with frequency and reaches the maximum value for the frequency close to Josephson plasma frequency. The larger angle $\varphi$, the larger the value of the maximum. At angles $\varphi$ larger than a certain critical value and at frequencies close to the Josephson plasma frequency, the dependences $\Pi_{\rm OSW}(\Omega)$ have divergences. The closer the azimuth angle to $90^o$, the closer the critical frequency $\Omega_{\rm cr}$  to one.

\section{Conclusion}

In this paper, we have theoretically examined the problem of the transition radiation by an electron moving along the normal to the interface between an isotropic dielectric and a layered superconductor. We have considered the geometry where the orientation of the superconducting layers is perpendicular to the interface. The analytical expressions for the strongly-anisotropic radiation fields and for the spectral density of the radiation energy flux have been derived.  We have found that the electron transit from a dielectric to a layered superconductor results not only in the generation of bulk Josephson plasma waves but also in the excitation of oblique surface electromagnetic waves. We show that the OSWs with frequencies close to the Josephson plasma frequency $\omega_J$ provide the main contribution to the OSWs energy flux. Moreover, for frequencies close to the Josephson plasma frequency, the spectral density of the OSWs radiation diverges at some critical value of the azimuth angle $\varphi$ with respect to the crystallographic \textbf{c}-axis. We have also studied the Cherenkov radiation by the electron escaping from the layered superconductor and show that this radiation is almost monochromatic. A remarkable feature of the Cherenkov radiation in a layered superconductor is that, contrary to the isotropic case, the Cherenkov radiation distinctly manifests itself in the angular dependence of the radiation energy flux.

Here we describe the electromagnetic properties of the layered superconductor by the effective permittivity tensor Eqs. (4--5). A similar anisotropic permittivity tensor may appear for other types of conducting lattices, e.g., in regular semiconductor heterostructures. Nevertheless, we emphasize that the above-mentioned effects are caused by the \emph{very strong} anisotropy of the current-carrying capability and are specific for layered superconductors.

\section{Acknowledgements}

We gratefully acknowledge partial support from the RIKEN iTHES project, MURI center for Dynamic Magneto-optics, JSPS-RFBR Contract No.~12-02-92100, Grant-in-Aid for Scientific Research (S), MEXT Kakenhi on Quantum Cybernetics, the JSPS-FIRST program, Ukrainian State Program on Nanotechnology, and the Program FPNNN of the NAS of Ukraine (grant No~0110U005642).

\appendix
\section{The electron's electromagnetic field components in the superconductor}

The electron's field components in the superconductor are,
\begin{equation}
   E^{({\rm el})}_{s \,x} (\vec{\kappa},\omega,y)=-\displaystyle\frac
   {i e k_x \Lambda_1 }{2\pi^2 v\Delta^{({\rm ord})}\Delta^{({\rm ext})}}
  \exp\left(i \frac{\omega}{v} y\right),\
\end{equation}
\begin{equation}
   E^{({\rm el})}_{s \,y} (\vec{\kappa},\omega,y)=-\displaystyle\frac
   {i e \Lambda_2}{2\pi^2\omega \Delta^{({\rm ord})}\Delta^{({\rm ext})}}\exp\left(i \frac{\omega}{v} y\right),\
\end{equation}
\begin{equation}
   E^{({\rm el})}_{s \,z} (\vec{\kappa},\omega,y)=-\displaystyle\frac
   {i e k_z}{2\pi^2 v\Delta^{({\rm ext})}}
   \exp\left(i \frac{\omega}{v} y\right),\
\end{equation}
\begin{equation}
    \vec{H}^{({\rm el})}_{s} (\vec{\kappa},\omega,y)=\displaystyle\frac{1}{c}\Bigl[\vec{v}\times
    \vec{D}^{({\rm el})}_{s}(\vec{\kappa},\omega, y)\Bigr],\
\end{equation}
where $D_{s\,x}^{({\rm el})}(\vec{\kappa},\omega, y)=\varepsilon_{ab}(\omega) E_{s\,x}^{({\rm el})}(\vec{\kappa},\omega,y)$, $D_{s\,y}^{({\rm el})}(\vec{\kappa},\omega, y)=\varepsilon_{ab}(\omega) E_{s\,y}^{({\rm el})}(\vec{\kappa},\omega,y)$, $D_{s\,z}^{({\rm el})}(\vec{\kappa},\omega, y)=\varepsilon_{c}(\omega) E_{s\,z}^{({\rm el})}(\vec{\kappa},\omega,y)$,
\begin{equation}
    \Lambda_1=k^2-\displaystyle\frac{\omega^2}{c^2}\varepsilon_c,\
\end{equation}
\begin{equation}
    \Lambda_2=\Bigl(k_{y}^2-\displaystyle\frac{\omega^2}{c^2}\varepsilon_{ab}\Bigr)\Bigl(k^2-\displaystyle\frac{\omega^2}{c^2}
\varepsilon_c\Bigr)-
 k_z^2 \displaystyle\frac{\omega^2}{c^2}\bigl(\varepsilon_c-\varepsilon_{ab}\bigr),\
\end{equation}
\begin{equation}
    \Delta^{({\rm ord})}=k^2-\displaystyle\frac{\omega^2}{c^2}\varepsilon_{ab},\
\end{equation}
\begin{equation}
\Delta^{({\rm ext})}=k_z^2 \varepsilon_c+\bigl(k_x^2+k_{y}^2\bigr) \varepsilon_{ab}-\displaystyle\frac{\omega^2}{c^2}\varepsilon_{ab}\varepsilon_{c}.
\end{equation}

\section{The field components for the ordinary and extraordinary waves}

From the Maxwell equations, we obtain the relationships between the Fourier components of the electric and magnetic fields in the ordinary and extraordinary waves.

For the ordinary wave, all field components can be expressed via the amplitude  $E^{(\rm ord)}_{d,s\,x}$ of the $x$-component of the electric field. These expressions for the dielectric region are,

\begin{equation}
    E^{(\rm ord)}_{d\,y}(\vec{\kappa},\omega,y)=-\displaystyle\frac{k_x}{k_{d\,y}}E^{(\rm ord)}_{d\,x}(\vec{\kappa},\omega,y),\ \label{appa}
\end{equation}
\begin{equation}
    H^{(\rm ord)}_{d\,x}(\vec{\kappa},\omega,y)=\displaystyle\frac{c k_x k_z}{\omega k_{d\,y}}E^{(\rm ord)}_{d\,x}(\vec{\kappa},\omega,y),\ \label{appb}
\end{equation}
\begin{equation}
    H^{(\rm ord)}_{d\,y}(\vec{\kappa},\omega,y)=\displaystyle\frac{c k_z}{\omega}E^{(\rm ord)}_{d\,x}(\vec{\kappa},\omega,y),\
\end{equation}
\begin{equation}
    H^{(\rm ord)}_{d\,z}(\vec{\kappa},\omega,y)=-\displaystyle\frac{c}{\omega}\frac{k_x^2+k_{d\,y}^2}{k_{d\,y}}E^{(\rm ord)}_{d\,x}(\vec{\kappa},\omega,y).
\end{equation}
The ordinary wave propagates in the dielectric with the $y$-component
  \begin{equation}
k_{d\,y}=-\sqrt{(\omega/c)^2\varepsilon_d-\kappa^2}
\end{equation}
of the wave vector, i.e., all the amplitudes in Eqs.~(B1)-(B4) are proportional to $\exp(ik_{d\, y} y)$.

For the superconducting region, we have
\begin{equation}
    E^{(\rm ord)}_{s\,y}(\vec{\kappa},\omega,y)=
    -\displaystyle\frac{k_x}{k^{(\rm ord)}_{s\,y}}E^{(\rm ord)}_{s\,x}(\vec{\kappa},\omega,y),\
\end{equation}
\begin{equation}
    H^{(\rm ord)}_{s\,x}(\vec{\kappa},\omega,y)=\displaystyle\frac{c k_x k_z}{\omega k^{(\rm ord)}_{s\,y}}E^{(\rm ord)}_{s\,x}(\vec{\kappa},\omega,y),\
\end{equation}
\begin{equation}
    H^{(\rm ord)}_{s\,y}(\vec{\kappa},\omega,y)=\displaystyle\frac{c k_z}{\omega}E^{(\rm ord)}_{s\,x}(\vec{\kappa},\omega,y),\
\end{equation}
\begin{eqnarray}
 &   H^{(\rm ord)}_{s\,z}(\vec{\kappa},\omega,y)=
    -\displaystyle\frac{c}{\omega}\frac{k_x^2+\bigl(k^{(\rm ord)}_{s\,y}\bigr)^2}{k^{(\rm ord)}_{s\,y}}\nonumber\\
 &\times
    E^{(\rm ord)}_{s\,x}(\vec{\kappa},\omega,y).
\end{eqnarray}
The $y$-component of the wave vector for the ordinary wave in the superconductor is
\begin{equation}
    k^{(\rm ord)}_{s\,y}=\sqrt{\displaystyle\frac{\omega^2}{c^2}\varepsilon_{ab}(\omega)-\kappa^2}.
\end{equation}

For the extraordinary wave, we obtain the following relationships:
\begin{equation}
    E^{({\rm ext})}_{d\,y}(\vec{\kappa},\omega,y)=\displaystyle\frac{k_{d\,y}}{k_x}E^{({\rm ext})}_{d\,x}(\vec{\kappa},\omega,y),\
\end{equation}
\begin{equation}
    E^{({\rm ext})}_{d\,z}(\vec{\kappa},\omega,y)=-\displaystyle\frac{k_x^2+k_{d\,y}^2}{k_x k_z}E^{({\rm ext})}_{d\,x}(\vec{\kappa},\omega,y),\
\end{equation}
\begin{equation}
    H^{({\rm ext})}_{d\,x}(\vec{\kappa},\omega,y)=-\displaystyle\frac{\omega}{c}\frac{k_{d\,y}}{k_x k_z}\varepsilon_d E^{({\rm ext})}_{d\,x}(\vec{\kappa},\omega,y),\
\end{equation}
\begin{equation}
    H^{({\rm ext})}_{d\,y}(\vec{\kappa},\omega,y)=\displaystyle\frac{\omega}{c k_z}\varepsilon_d E^{({\rm ext})}_{d\,x}(\vec{\kappa},\omega,y)
\end{equation}
 in the dielectric and
\begin{equation}
    E^{({\rm ext})}_{s\,y}(\vec{\kappa},\omega,y)=
    \displaystyle\frac{k_{s\,y}^{({\rm ext})}}{k_x}E^{({\rm ext})}_{s\,x}(\vec{\kappa},\omega,y),\
\end{equation}
    \begin{eqnarray}
 &   E^{({\rm ext})}_{s\,z}(\vec{\kappa},\omega,y)=-\displaystyle\frac{\varepsilon_{ab}(\omega)}
    {\varepsilon_c(\omega)}\displaystyle\frac{k_x^2+\bigl(k_{s\,y}^{({\rm ext})}\bigr)^2}{k_x k_z}\nonumber\\
 &   \times
    E^{({\rm ext})}_{s\,x}(\vec{\kappa},\omega,y),\
    \end{eqnarray}
\begin{eqnarray}
 &   H^{({\rm ext})}_{s\,x}(\vec{\kappa},\omega,y)=-\displaystyle\frac{\omega}{c}\frac{k_{s\,y}^{({\rm ext})}}{k_x k_z}\varepsilon_{ab}(\omega)\nonumber\\
 &   \times
    E^{({\rm ext})}_{s\,x}(\vec{\kappa},\omega,y),\
    \end{eqnarray}
\begin{equation}
    H^{({\rm ext})}_{s\,y}(\vec{\kappa},\omega,y)=\displaystyle\frac{\omega}{c k_z}\varepsilon_{ab}(\omega)E^{({\rm ext})}_{s\,x}(\vec{\kappa},\omega,y),\
\end{equation}
\begin{equation}
    k^{({\rm ext})}_{s\,y}= \sqrt{\displaystyle\frac{\omega^2}{c^2}\varepsilon_{c}(\omega)-\frac{\varepsilon_c(\omega)}{\varepsilon_{ab}(\omega)}k_z^2-k_x^2}
\end{equation}
in the superconductor.

\section{Total radiation field in the dielectric}

Omitting the common multiplier $\exp[i(\vec{\kappa}{\vec \rho} + k_{d\,y} y -\omega t)]$, we can present the expressions for all components of the total radiation electromagnetic field in the form,
\begin{equation}
    E^{\rm rad}_{d\,x}(\vec{\kappa},\omega)=-\displaystyle\frac{i\omega}{c}\frac{\alpha_1 Q_2+\alpha_3 Q_1}{\Delta_{\rm OSW}},\
\end{equation}
\begin{eqnarray}
 &   E^{\rm rad}_{d\,y}(\vec{\kappa},\omega)=\displaystyle\frac{i\omega}{c k_{d\,y}\Delta_{\rm OSW}}\nonumber\\
 &  \times \Bigl[\bigl(\alpha_3 k_x+\alpha_2 k_z\bigr) Q_1+\bigl(\alpha_1 k_x-\alpha_4 k_z\bigr) Q_2\Bigr],\
\end{eqnarray}
\begin{equation}
    E^{\rm rad}_{d\,z}(\vec{\kappa},\omega)=\displaystyle\frac{i\omega}{c}\frac{\alpha_4 Q_2-\alpha_2 Q_1}{\Delta_{\rm OSW}},\
\end{equation}
\begin{eqnarray}
 &    H^{\rm rad}_{d\,x}(\vec{\kappa},\omega)=\displaystyle\frac{c}{\alpha_1 k_{d\,y}\omega}  E_{d\,x}(\vec{\kappa},\omega) \nonumber\\
 &   \times \Bigl[\alpha_1 k_x k_z - \alpha_4 \bigl(k_z^2+k_{d\,y}^2\bigr)\Bigr]-i\displaystyle\frac{k_z^2+k_{d\,y}^2}{\alpha_1 k_{d\,y}} Q_1,\
\end{eqnarray}
\begin{equation}
    H^{\rm rad}_{d\,y}(\vec{\kappa},\omega)=\displaystyle\frac{c}{\omega}\Bigl(k_z+\displaystyle\frac{\alpha_4}{\alpha_1} k_x\Bigr) E_{d\,x}(\vec{\kappa},\omega)
    +i\displaystyle\frac{k_x}{\alpha_1}Q_1,\
\end{equation}
\begin{eqnarray}
 &   H^{\rm rad}_{d\,z}(\vec{\kappa},\omega)=-\displaystyle\frac{c}{\alpha_1 k_{d\,y} \omega}E_{d\,x}(\vec{\kappa},\omega)\nonumber\\
 &  \times \Bigl[\alpha_1\bigl(k_x^2+k_{d\,y}^2\bigr)-\alpha_4 k_x k_z\Bigr]+i \displaystyle\frac{k_x k_z}{\alpha_1 k_{d\,y}} Q_1
\end{eqnarray}
where
\begin{equation}
    \alpha_1=-k_x k_z \displaystyle\frac{k_{s\,y}^{({\rm ord})}-k_{d\,y}}{k_{d\,y}k_{s\,y}^{({\rm ord})}},\
     \alpha_2=-k_x k_z [\varepsilon_{ab}(\omega)-\varepsilon_d],\
\end{equation}
\begin{equation}
    \alpha_3=k_z^2 [\varepsilon_{ab}(\omega)-\varepsilon_d]+\varepsilon_{ab}(\omega) k_{d\,y} (k_{d\,y}-k_{s\,y}^{({\rm ext})}),\
\end{equation}
\begin{equation}
    \alpha_4=\displaystyle\frac{1}{k_{s\,y}^{({\rm ord})}}\Bigl[\displaystyle\frac{\omega^2}{c^2}\varepsilon_{ab}(\omega)-k_z^2\Bigr]
-\displaystyle\frac{1}{k_{d\,y}}\Bigl(\displaystyle\frac{\omega^2}{c^2}\varepsilon_d-k_z^2\Bigr),\
\end{equation}
\begin{equation}
    \Delta_{\rm OSW}=\alpha_1 \alpha_2 + \alpha_3 \alpha_4,\
\end{equation}
\begin{eqnarray}
 &   Q_1=-\displaystyle\frac{e c k_x}{2\pi^2 v \omega\varepsilon_d k_{s\,y}^{({\rm ord})} \Delta_1 \Delta^{({\rm ord})}
    \Delta^{({\rm ext})}}\nonumber\\
 & \times  \Biggl\{ \Bigl[\Delta^{({\rm ord})}\Delta^{({\rm ext})}-\varepsilon_d \Delta_1 \bigl(k^2-\displaystyle\frac{\omega^2}{c^2}\varepsilon_c(\omega)\bigr)\Bigr]\Bigl[\displaystyle\frac{\omega^2}{c^2}\varepsilon_{ab}(\omega)-k_z^2\Bigr]\nonumber\\
  & -\displaystyle\frac{ \omega v}{c^2} k_{s\,y}^{({\rm ord})}\varepsilon_d
  \Bigl[\Delta^{({\rm ord})}\Delta^{({\rm ext})}-\varepsilon_{ab}(\omega) \Delta_1 \bigl(k^2-\displaystyle\frac{\omega^2}{c^2}\varepsilon_c(\omega)\bigr)\Bigr]\nonumber\\
&  +k_z^2 \Delta^{({\rm ord})} (\Delta^{({\rm ext})}-\varepsilon_d \Delta_1)\Biggr\},\
\end{eqnarray}
\begin{eqnarray}
 &   Q_2=\displaystyle\frac{e c k_z k_{d\,y}}{2\pi^2 v \omega^2 \Delta_1 \Delta^{({\rm ord})}
    \Delta^{({\rm ext})}}\nonumber\\
 & \times  \Biggl\{\displaystyle v \Delta^{({\rm ord})} [\Delta^{({\rm ext})}-\varepsilon_c (\omega)\Delta_1 ]\Bigl[\displaystyle\frac{\omega^2}{c^2}\varepsilon_{ab}(\omega)-k_z^2\Bigr]\nonumber\\
  &- v k_x^2
  \Bigl[\Delta^{({\rm ord})}\Delta^{({\rm ext})}-\varepsilon_{ab}(\omega) \Delta_1 \bigl(k^2-\displaystyle\frac{\omega^2}{c^2}\varepsilon_c(\omega)\bigr)\Bigr]\nonumber\\
&  -\displaystyle\frac{\omega\varepsilon_{ab}(\omega) }{\varepsilon_d}k_{s\,y}^{({\rm ext})} \Delta^{({\rm ord})} (\Delta^{({\rm ext})}-\varepsilon_d \Delta_1)\Biggr\},\
\end{eqnarray}
\begin{equation}
    \Delta_1=k^2-\displaystyle\frac{\omega^2}{c^2}\varepsilon_d,\
    k^2=\kappa^2+\displaystyle\frac{\omega^2}{v^2}.
\end{equation}

\section{Radiation field of the oblique surface waves}

In order to derive the spatial and temporal dependence of the OSWs fields in an explicit form, we need to integrate Eqs.~(C1--C6) over  $k_x$, $k_z$,  and  $\omega$, taking into account the poles of the integrands. As a result, we obtain the following expressions:
\begin{eqnarray}
&    E_{d\,x}^{\rm sw}(\vec{r},t)=-\displaystyle\frac{\pi}{c}\int\limits_0^{2\pi}d\psi\int\limits_{-\infty}^{\infty}d\omega
\displaystyle\frac{\omega (\alpha_1 Q_2+\alpha_3 Q_1)}{\Gamma(\omega,\psi)}\nonumber\\
&\times\exp\bigl\{i\bigl[\rho \kappa_0 \cos(\psi-\varphi)+k_{d\,y}y-
\omega t\bigr]\bigr\},
\end{eqnarray}
\begin{eqnarray}
&    E_{d\,y}^{\rm sw}(\vec{r},t)=\displaystyle\frac{\pi}{c} \int\limits_0^{2\pi}d\psi\int\limits_{-\infty}^{\infty}d\omega
\displaystyle\frac{\omega}{k_{d\,y}\Gamma(\omega,\psi)}\nonumber\\
&\times \bigl[(k_x\alpha_3+k_z \alpha_2) Q_1+ (k_x\alpha_1-k_z \alpha_4) Q_2)\bigr] \nonumber\\
&\times\exp\bigl\{i\bigl[\rho \kappa_0 \cos(\psi-\varphi)+k_{d\,y}y-
\omega t\bigr]\bigr\},
\end{eqnarray}
\begin{eqnarray}
&    E_{d\,z}^{\rm sw}(\vec{r},t)=\displaystyle\frac{\pi}{c}\int\limits_0^{2\pi}d\psi\int\limits_{-\infty}^{\infty}d\omega
\displaystyle\frac{\omega (\alpha_4 Q_2-\alpha_2 Q_1)}{\Gamma(\omega,\psi)}\nonumber\\
&\times\exp\bigl\{i\bigl[\rho \kappa_0 \cos(\psi-\varphi)+k_{d\,y}y-
\omega t\bigr]\bigr\},
\end{eqnarray}
\begin{eqnarray}
&    H_{d\,x}^{\rm sw}(\vec{r},t)=\pi\int\limits_0^{2\pi}d\psi\int\limits_{-\infty}^{\infty}d\omega
\displaystyle\frac{(\alpha_1 Q_2+\alpha_3 Q_1)}{\alpha_1 k_{d\,y}\Gamma(\omega,\psi)}\nonumber\\
&\times\bigl[\alpha_1 k_x k_z - \alpha_4 (k_z^2+k_{d\,y}^2)\bigr]\nonumber\\
&\times\exp\bigl\{i\bigl[\rho \kappa_0 \cos(\psi-\varphi)+k_{d\,y}y-
\omega t\bigr]\bigr\},
\end{eqnarray}
\begin{eqnarray}
&    H_{d\,y}^{\rm sw}(\vec{r},t)=\pi\int\limits_0^{2\pi}d\psi\int\limits_{-\infty}^{\infty}d\omega
\displaystyle\frac{(\alpha_1 Q_2+\alpha_3 Q_1)}{\Gamma(\omega,\psi)}\nonumber\\
&\times\Bigl(k_z + k_x \displaystyle\frac{\alpha_4}{\alpha_1}\Bigr)\nonumber\\
&\times\exp\bigl\{i\bigl[\rho \kappa_0 \cos(\psi-\varphi)+k_{d\,y}y-
\omega t\bigr]\bigr\},
\end{eqnarray}
\begin{eqnarray}
&    H_{d\,z}^{sw}(\vec{r},t)=-\pi\int\limits_0^{2\pi}d\psi\int\limits_{-\infty}^{\infty}d\omega
\displaystyle\frac{(\alpha_1 Q_2+\alpha_3 Q_1)}{\alpha_1 k_{d\,y}\Gamma(\omega,\psi)}\nonumber\\
&\times\bigl[\alpha_1 (k_x^2+k_{d\,y}^2) - \alpha_4 k_x k_z\bigr]\nonumber\\
&\times\exp\bigl\{i\bigl[\rho \kappa_0 \cos(\psi-\varphi)+k_{d\,y}y-
\omega t\bigr]\bigr\},
\end{eqnarray}
\begin{eqnarray}
 & \Gamma(\omega,\psi)=\displaystyle\frac{k_{d\,y}-k_{s\,y}^{({\rm ord})}}{k_{2y}^{({\rm ord})}}\Biggl[
 \displaystyle\frac{\kappa^2-2 k_{s\,y}^{({\rm ord}) 2}}{k_{s\,y}^{({\rm ord})}}[\varepsilon_{ab}(\omega)-\varepsilon_d]\cos^2\psi\nonumber\\
 &+\varepsilon_{ab}(\omega)\left(\displaystyle\frac{1}{k_{d\,y}}-\displaystyle\frac{Q_3}{k_{2y}^{({\rm ext})}}\right)\Bigl(k_{d\,y} k_{2y}^{({\rm ord})}-\kappa^2\sin^2\psi\Bigr)
 \nonumber\\
 &+ \varepsilon_{ab} \left(k_{d\,y}-k_{s\,y}^{({\rm ext})}\right)\left(\displaystyle\frac{k_{s\,y}^{({\rm ord})}}{k_{d\,y}}+\displaystyle\frac{k_{d\,y}}{k_{s\,y}^{({\rm ord})}}+2\sin^2\psi\right)
 \Biggr],
\end{eqnarray}
where $k_x=\kappa_0(\omega,\psi)\sin\psi$, $k_z=\kappa_0(\omega,\psi)\cos\psi$, $\kappa_0(\omega,\psi)$ is the solution of the dispersion relation $\Delta_{\rm OSW}=0$, $\psi$ is the angle between the wave vector $\vec{\kappa}$ and the crystallographic $\bf{c}$-axis, and $Q_3=(\varepsilon_c/\varepsilon_{ab})\cos^2\psi+\sin^2\psi$. Recall that $\varphi$ is the azimuth angle between the radius vector $\vec{\rho}$ and the crystallographic  $\bf{c}$-axis (see Fig.~\ref{Fig1}).

\end{document}